\documentstyle[twoside,fleqn,espcrc2,epsfig,xspace]{article}

\begin{document}

%\draft
%\preprint{HEP/123-qed}

\title{New Limit on Muon and Electron Lepton Number Violation from $K^0_L \rightarrow \mu^\pm e^\mp$ Decay}

\author{
D. Ambrose$^{(a)}$,
C. Arroyo$^{(b)}$,
M. Bachman$^{(c)}$,
D. Connor$^{(c)}$,
M. Eckhause$^{(d)}$,
S. Graessle$^{(a)}$,
A. D. Hancock$^{(d)}$,
K. Hartman$^{(b)}$,
M. Hebert$^{(b)}$,
C. H. Hoff$^{(d)}$,
G. W. Hoffmann$^{(a)}$,
G. M. Irwin$^{(b)}$,
J. R. Kane$^{(d)}$,
N. Kanematsu$^{(c)}$,
Y. Kuang$^{(d)}$,
K. Lang$^{(a)}$,
R. Lee$^{(c)}$,
R. D. Martin$^{(d)}$,
J. McDonough$^{(a)}$,
A. Milder$^{(a)}$,
W. R. Molzon$^{(c)}$,
M. Pommot-Maia$^{(b)}$,
P. J. Riley$^{(a)}$,
J. L. Ritchie$^{(a)}$,
P. D. Rubin$^{(e)}$,
V. I. Vassilakopoulos$^{(a)}$,
R. E. Welsh$^{(d)}$,
S. G. Wojcicki$^{(b)}$\\
(BNL E871 Collaboration)\\
   $^{(a)}$University of Texas, Austin, Texas, 78712,\\
   $^{(b)}$Stanford University, Stanford, California, 94305,\\
   $^{(c)}$University of California, Irvine, California, 92697,\\ 
   $^{(d)}$College of William and Mary, Williamsburg, Virginia, 23187, \\
   $^{(e)}$University of Richmond, Richmond, Virginia, 23173\\ 
}

%\date{\today}
%\maketitle

\begin{abstract}
The most sensitive experiment to date to search for the muon and electron
lepton
number violating decay $K^0_L \rightarrow \mu^\pm e^\mp$ has detected no
events consistent with this process.  Based on this result, the
90\% confidence level upper limit on the branching fraction is 
$B(K^0_L \rightarrow \mu^\pm e^\mp) < 4.7\times10^{-12}$.
\end{abstract}

\maketitle

%\pacs{PACS numbers: 13.20.Eb, 11.30.Hv}

%\narrowtext

This Letter reports a significantly improved experimental upper limit
on the rate for the decay $K^0_L \rightarrow \mu^\pm e^\mp$.
This process would violate the conservation of muon and electron lepton
number (referred to as 
separate lepton number) while
conserving total lepton number; it is not allowed in the Standard Model of
particle physics. Incorporating neutrino masses and mixing into the Standard
Model, consistent with current information on these quantities,
leads to predicted rates for $K^0_L \rightarrow \mu^\pm e^\mp$ well  
below experimental sensitivities~\cite{Langacker:1988}. Hence, 
observation of this decay would signal new physics processes. 

Experiments to search directly for separate lepton number violation
have been performed for many years, all with null results. Some of the
best limits come from previous searches for
$K_L^0 \rightarrow \mu^{\pm} e^{\mp}$~\cite{Arisaka:1993,Akagi:1995},
for which the combined upper limit is $2.4\times10^{-11}$,
and from searches for
$K_L^0 \rightarrow \pi^0 \mu^{\pm}  e^{\mp}$~\cite{Arisaka:1998},
$K^+ \rightarrow \pi^+ \mu^+  e^-$~\cite{Lee:1990}, 
$\mu^+ \rightarrow e^+ \gamma$~\cite{Bolton:1988}, 
$\mu^+ \rightarrow e^+e^+e^-$~\cite{Bellgardt:1988}, 
and $\mu^- N \rightarrow e^- N$~\cite{Dohmen:1993}.
The sensitivity of these processes to mechanisms which allow for
separate lepton number violation varies.
Several theoretical models allow $K^0_L \rightarrow \mu^\pm e^\mp$,
including some at rates as large as current experimental limits:
horizontal gauge interactions~\cite{Cahn:1980}, left-right
symmetry~\cite{Langacker:1988,Gagyi:1998},
technicolor~\cite{Dimopoulos:1981}, compositeness~\cite{Pati:1986},
and supersymmetry~\cite{Mukhopadhyaya:1990}. 

The experiment (E871) was performed in the B5 beam line of the
Alternating Gradient
Synchrotron at Brookhaven National Laboratory~(BNL). 
Data were collected during running periods in 1995 and 1996.
A 24 GeV proton beam incident on a 1.4 interaction length platinum
target produced the neutral beam.  A targeting angle of
3.75$^{\circ}$ was chosen to maximize the $K_L^0$ yield
while minimizing the $n/K_L^0$ ratio.
Sweeping magnets downstream of the
target removed charged particles. Thin lead foils in the
first magnet aperture reduced the $\gamma$ flux in the beam.
Three collimators defined the neutral beam to subtend
a solid angle of approximately 4$\times$16~mrad$^2$.
The proton intensity was typically
$1.5 \times 10^{13}$ in a 1.2-1.6~s pulse every 3.2-3.6~s, resulting
in $2\times10^8$ $K^0_L$ per pulse ($2<p_K<16$~GeV/c), 7.5\% of which decayed
in an evacuated decay volume between 9.75 and 20.75~m from the target. 
We measured the $n / K^0_L$ ratio to be $8\pm3$.

\begin{figure*}
\begin{center}
\epsfxsize=13.5cm
\epsfig{figure=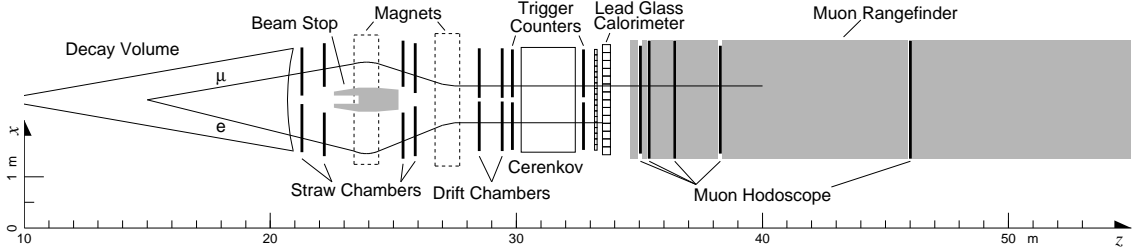,width=160mm}
\caption{Plan view of the E871 beam line and apparatus.}
\label{fig:apparatus}
\end{center}
\end{figure*}

The experimental apparatus, illustrated in Fig.~\ref{fig:apparatus}, has
been described previously~\cite{Ambrose:1998}.  Accordingly, we describe
here only the most important features, emphasizing characteristics of 
the detector most germane to this search.
A magnetic spectrometer consisting of six pairs
of tracking chambers and two magnets was used to measure kinematic parameters
of charged $K_L^0$ decay products.  A beam stop~\cite{Belz:1998}
was placed in the first magnet to 
absorb the neutral beam. The spectrometer was followed by 
scintillation hodoscopes (TSC) at two longitudinal positions,
used to select events with two charged particles and define the event time.
Both hodoscopes contained one $x$~(horizontal) measuring plane on either
side of the detector.  The downstream module contained an additional
plane on either side providing measurements in the $y$~(vertical) view.
A segmented threshold
$\check{\mathrm{C}}$erenkov detector (CER) and lead glass calorimeter
(PBG) were used to identify electrons.
A 30.5 cm thick iron filter followed the PBG.
Muons were identified by scintillation hodoscopes (MHO) and a range finder
(MRG) downstream of it. Six MHO planes at five different $z$ locations 
provided position and timing information.  The MRG consisted of
26 vertical and 26 horizontal planes of proportional tube hodoscopes,
located between absorbers of steel, marble, and aluminum.
The amount of material between successive planes increased with depth
and corresponded to 5\% increments of muon range.

The lowest level trigger (L0) required a coincidence of signals in all
six TSC planes, providing two $x$ and one $y$
measurements on each side of the detector.
Further, a parallelism condition was imposed: each pair of $x$
view signals was required to be consistent with a 
particle trajectory with $|dx/dz| < 31$~mrad.
This requirement maintained good acceptance for two-body decays
while reducing the acceptance for the dominant three-body modes.
The typical L0 trigger rate was approximately
70~kHz.

The next trigger level (L1) was
designed to accept all dilepton decay modes; it was formed from
a coincidence of an L0 trigger and signals from 
particle identification (PID) detectors.
The muon signal was taken from the MHO plane located at a $z$ position 
corresponding to an energy loss of 1~GeV.
The electron signal was provided by the $\check{\mathrm{C}}$erenkov detector. 
Spatial correlation between TSC and PID signals was required.
In addition to the dilepton modes, one of every 1000 L0 trigger events
was selected to form the ``minimum 
bias'' sample. It was used for detector calibration and for flux
determination, based on the number of $K^0_L \rightarrow \pi^+ \pi^-$ events
in that sample.  The L1 trigger rate averaged about 7~kHz.

Events satisfying the L1 trigger were digitized and
transferred to a set of eight processors in which a software trigger (L3) was
implemented. It did fast event
reconstruction using hits in the TSC and all tracking detectors.  At least one
track on each side of the spectrometer and a decay vertex 
within the neutral
beam were required.  In addition, a two-body invariant mass $M_{+-}$
exceeding 460~MeV/c$^2$
and two-body transverse momentum $p_T$ less than 60~MeV/c were required
for the $\mu e$ trigger.
Charged particle mass assignments were determined by
the triggering PID detector. An event with more than one L1 dilepton
hypothesis was accepted if any of the triggered modes satisfied the L3
criteria. The L3 trigger algorithm was run on minimum bias
events but no selection on kinematic quantities was made.
All events passing the L3
trigger were written to tape for off-line processing.

The off-line pattern recognition software used a similar but more thorough
algorithm for pattern recognition. More rigorous selection criteria
were applied 
to geometric and kinematic quantities. Events were selected if $M_{+-}$
exceeded 470~MeV/c$^2$ and if either of the requirements 
$p_T < 40$~MeV/c or $\theta_c < 4.5$~mrad was satisfied, where
$\theta_c$ is the angle between the kaon direction (determined from the 
target and vertex position) 
and the direction of the momentum sum vector of the two charged particles.

The $K^0_L \rightarrow \mu^\pm e^\mp$ candidate sample was selected
from events satisfying these criteria and having a $\mu e$ L1 hypothesis.
Projected track positions were required to be spatially consistent
with signals in the PID detectors.
A particle was identified as an electron if it
had appropriate signals in the CER and PBG, the track time determined from the 
tracking spectrometer was consistent with the
CER time, and the momentum was consistent with the energy deposited
in the PBG.  A particle was identified as a muon if it projected to
hits in the MHO and MRG, the track time was consistent with the
MHO signal times, and the energy inferred from the MRG exceeded 80\% of the
momentum.
Also, any PBG signal associated with a muon candidate was required
to be inconsistent with that of an electron.

The best estimates of kinematic parameters of all events satisfying
the above criteria were determined using two independent algorithms
that had different sensitivities to errors in pattern recognition.
One algorithm (FT) minimized a $\chi^2$ for a kinematic fit  to all hits in
the tracking detectors, appropriately accounting for  detector resolution
and multiple scattering using an error matrix.  
The second algorithm (QT) did separate calculations of kinematic quantities
for the front and back halves of the spectrometer and used the two
measurements to form a single momentum and a measure of the track quality.
The FT algorithm had a better mass resolution 
(1.13 versus 1.26~MeV/c$^2$ for QT
for $K_L^0 \rightarrow \pi^+ \pi^-$ events) and was used to determine
the event kinematics.  It was required that $M_{\mu e}$ and $p_T$ 
determined by the fitters be consistent.

Events were required to be consistent with originating from a single kaon
decay.  The times of the two particles as measured by the spectrometer
were required to be consistent within measurement uncertainty and
their trajectories were required to project to a common vertex
with $z > 9.75$ m and transverse coordinates within the neutral beam.

Events were selected on the basis of the quality of the kinematic fit in
order to reject those with 
tracking mismeasurement or with pion decay or large angle 
scattering within the spectrometer.
In particular, the measurements of
 front and  back momenta were required to be consistent, and events 
with poor tracking $\chi^2$ were rejected.
High momentum particles were particularly susceptible to errors
in track reconstruction, and a maximum 
momentum of 8~GeV/c was imposed. This requirement also ensured that
muons were contained
in the range-finder and that $\check{\mathrm{C}}$erenkov signals were not
caused by pions.
Events were rejected if particle trajectories  projected to any material 
in the spectrometer volume other than the relevant detectors.

To ensure that event selection criteria were free of bias from knowledge
of potential signal events, they were chosen by studying
$K^0_L \rightarrow \mu^\pm e^\mp$ candidates with $M_{\mu e}>485$~MeV/c$^2$
and $p_T^2<900$~(MeV/c)$^2$, but excluding a region,
$490 < M_{\mu e} < 505$~MeV/c$^2$ and $p_T^2 < 100$~(MeV/c)$^2$,
larger than the potential $K^0_L \rightarrow \mu^\pm e^\mp$ signal region.
This exclusion region corresponded to a $\pm5\,\sigma$ interval in mass
and a 3$\,\sigma$ interval in $p_T^2$.  

The primary source of $K^0_L \rightarrow \mu^\pm e^\mp$ background is
$K_L^0 \rightarrow \pi e \nu$ in which a pion decays upstream of the muon
filter (about 4\% of all $K_L^0$ decays).  Misidentification of the
pion as a muon results in a maximum value for $M_{\mu e}$ of 489.3~MeV/c$^2$
if track momenta and directions are correctly measured and the pion is
assigned a muon mass.
The resolution in $M_{\mu e}$ inferred from the measured
$K^0_L \rightarrow \pi^+ \pi^-$ mass resolution was 1.38~MeV/c$^2$;
hence, background arising from Gaussian tails in the mass 
measurement is negligible, but non-Gaussian effects could be important.
One source of non-Gaussian errors is elastic scattering  
in the vacuum window or in the first tracking detector
(0.12\% and 0.23\% radiation lengths, respectively),
 referred to as upstream scatters.  
Scattering and decay in the plane normal to 
 $\vec{p}_e\times\vec{p}_{\pi}$ (the $K_L^0$ decay plane) 
may increase the $\pi e$ opening angle and hence the 
value of $M_{\mu e}$, while scattering or decay out of the plane cannot 
increase $M_{\mu e}$ significantly. 
Monte Carlo simulations showed that the dominant background occurs when a low 
energy electron elastically scatters upstream and the
pion decays upstream of the spectrometer. 

\begin{figure}
\begin{center}
\epsfig{figure=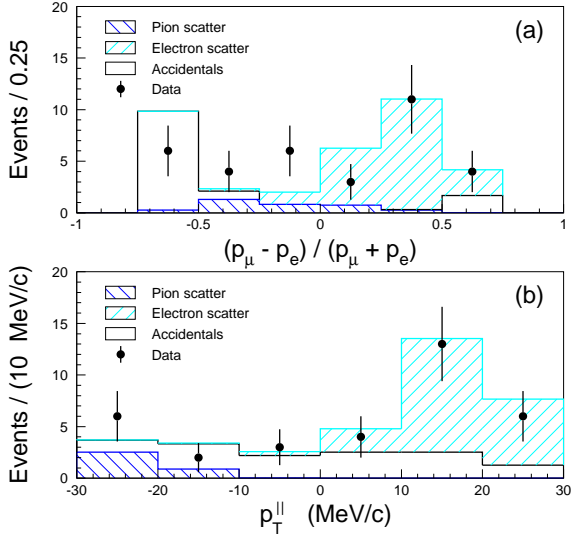,width=75mm}
\caption{Data and Monte Carlo distributions of (a) momentum asymmetry and
(b) $p_T^{^{_\parallel}}$.  Events shown satisfy all selection criteria
except those on the quantities displayed above and
have $M_{\mu e}>493$~MeV/c$^2$ and $100<p_T^2< 900$~(MeV/c)$^2$.
\label{fig:pasym_ptin}}
\end{center}
\end{figure}

To study this background in detail we imposed the requirement $p_e > 1$~GeV/c
and examined the remaining high mass events outside of the exclusion region.
The momentum asymmetry and the component of the $\mu e$ transverse momentum
in the decay plane (denoted by $p_T^{^{_\parallel}}$) are shown in
Fig.~\ref{fig:pasym_ptin}
together with the Monte Carlo predictions.  The sign of $p_T^{^{_\parallel}}$
is taken to be positive if it lies on the electron side of the spectrometer.
The upstream scatter events are characterized by large momentum asymmetry
and more importantly by large $p_T^{^{_\parallel}}$.
Without additional selection criteria, our expected
background in an appropriate signal region would be about one event.
We imposed the requirement that $(p_{\mu} - p_e)/(p_{\mu} + p_e)<0.5$
and that $p_T^{^{_\parallel}}$ be small~\cite{ptin}, which reduced this
background significantly.

A second potential source of background is accidental coincidences of
$K_L^0 \rightarrow \pi e \nu$ and $K_L^0 \rightarrow \pi \mu \nu$ decays.
Because the muon and electron originate from
independent decays, they can reconstruct with a value of $M_{\mu e} > M_K$.
Monte Carlo simulations were done
to study this background. In many of these events at least one of the pion
trajectories is fully contained in 
the spectrometer, and the background results when the $x$ and $y$ view 
tracks on the pion side are mispaired.
Rejecting events with three or more fully reconstructed
tracks in the spectrometer reduced this background by an order of
magnitude.

\begin{figure}
\begin{center}
\epsfig{figure=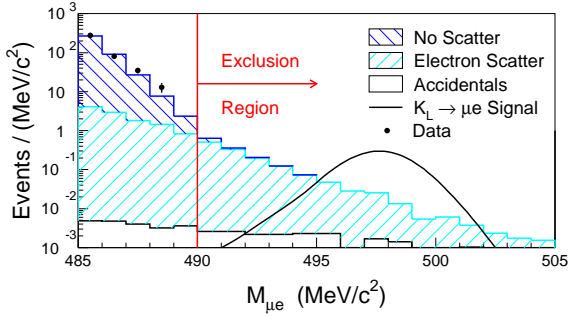,width=75mm}
\caption{Data and Monte Carlo distributions of $M_{\mu e}$
for events with $p_T^2 < 20$~(MeV/c)$^2$
(the $p_T^2$ range of the signal region).
We estimate the uncertainty in the Monte Carlo normalization to be 10\%.
The calculated $K_L^0 \rightarrow \mu^{\pm} e^{\mp}$ signal curve
assumes a branching fraction of $2.1\times10^{-12}$.
}
\label{fig:signal_mass}
\end{center}
\end{figure}

The $K_L^0 \rightarrow \mu^\pm e^\mp$ selection criteria,
including choice of the signal region, were determined by
simultaneously varying values of relevant event selection parameters
to maximize the sensitivity to signal while suppressing the expected
contribution from the dominant source of background to 0.1
event.
The calculated background after application of all selection criteria except
those on $M_{\mu e}$ and $p_T^2$ is
compared to the data in Fig.~\ref{fig:signal_mass}.
For $M_{\mu e}<490$~MeV/c$^2$,
the background is dominated by correctly measured $K_L^0 \rightarrow \pi e \nu$
decays.  For $M_{\mu e} > 493$~MeV/c$^2$, events passing the selection criteria
are dominated by large upstream scatters, with a lesser contribution from
accidentals.  Reducing the background by an additional factor of 10
would require tighter selection criteria which would result in a
50\% acceptance loss.

\begin{figure}
\begin{center}
\epsfig{figure=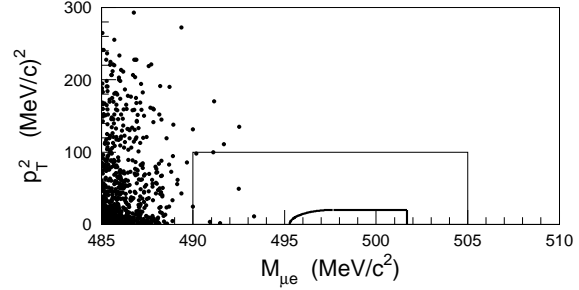,width=75mm}
\caption{Plot of $p_T^2$ versus $M_{\mu e}$.
The exclusion region for the blind analysis is indicated by the
box. The signal region is indicated by the smaller contour.
}
\label{fig:pt2_mass}
\end{center}
\end{figure}

For  $M_{\mu e} < M_K$, the
signal region was defined by the ellipse
$[\, p_T^2 / (20 \, ($MeV/c$)^2)]^2 + [\Delta M / (2.4$ MeV/c$^2)]^2 < 1$,
where $\Delta M = M_{\mu e} - M_K$.
For $M_{\mu e} > M_K$, the signal region
was defined by $\Delta M < 4$~MeV/c$^2$ and
$p_T^2 < 20 \ (\mathrm{MeV/c})^2$.
The background is larger in the region $M_{\mu e}<M_K$; hence
different shapes were chosen for the signal region below and above $M_K$
as a compromise between acceptance and background rejection.
After all selection criteria (including the choice of the
signal region) were determined, all data (including those in the exclusion
region) were reanalyzed. Figure~\ref{fig:pt2_mass} shows the final
distribution in $p_T^2$ versus $M_{\mu e}$.
There are no events in the signal region.

The $K^0_L \rightarrow \mu^\pm e^\mp$ sensitivity is determined from the
number of $K^0_L \rightarrow \pi^+ \pi^-$ decays in the minimum bias sample.
These events were required to satisfy an appropriate subset of the final
selection criteria discussed above, and were required to
have no PBG signals consistent with those of an electron.
A fit in the $p_T^2$ versus $M_{\pi \pi}$ plane 
was done to subtract residual $K_L^0 \rightarrow \pi \mu \nu$ background 
and to determine the number of $K^0_L \rightarrow \pi^+ \pi^-$ events.
Small differences in geometric acceptance and cut efficiencies were determined
by Monte Carlo simulation. 

In the absence of any signal events, the 90\% confidence level upper
limit on the $K^0_L \rightarrow \mu^\pm e^\mp$ branching fraction is given by 

{\setlength\arraycolsep{2pt}
\begin{eqnarray}
B(K_L^0 \rightarrow \mu^\pm e^\mp) & \!< & \!2.3 \, B_{\pi \pi} 
\frac{f_{\pi \pi}}{R N_{\pi \pi}} \frac{A_{\pi \pi}}{A_{\mu e}} 
\frac{1}{\epsilon^{\mathrm{L1}}_{\mu e}} 
\frac{1}{\epsilon^{\mathrm{L3}}_{\mu e}} 
\frac{\epsilon^{\mathrm{PID}}_{\pi \pi}}{\epsilon^{\mathrm{PID}}_{\mu e}}
\nonumber
\end{eqnarray}}

\noindent where $B_{\pi \pi}$~\cite{Caso:1998} is the
$K^0_L \rightarrow \pi^+ \pi^-$ branching fraction, $R$ is the $\pi \pi$
prescale (a hardware factor of 1000 times a software factor of
20), $N_{\pi \pi}$ is the number of $K^0_L \rightarrow \pi^+ \pi^-$ events
in the prescaled minimum bias sample (including a 0.05\% correction for
$K_L-K_S$ interference),
$f_{\pi \pi}$ is a factor to account for loss of $K^0_L \rightarrow \pi^+\pi^-$
events due to hadronic interactions in the spectrometer,
$A_{\pi \pi}$ and $A_{\mu e}$ are the mode dependent
acceptances (including geometric acceptance and selection criteria efficiency),
$\epsilon^{\mathrm{L1}}_{\mu e}$ and 
$\epsilon^{\mathrm{L3}}_{\mu e}$ are the efficiencies of the 
L1 and L3 triggers, and
$\epsilon^{\mathrm{PID}}_{\pi \pi}$ and $\epsilon^{\mathrm{PID}}_{\mu e}$
are the efficiencies of the particle identification.
The geometric acceptance for $K^0_L \rightarrow \mu^\pm e^\mp$
($K^0_L \rightarrow \pi^+ \pi^-$) decays with
$9.75<z<20.75$~m and kaon momentum $2<p_K<16$~GeV/c was
2.36\% (2.63\%).  The parallelism requirement reduced this
to 1.97\% (2.21\%).  Event selection criteria further reduced
the acceptance to 1.14\% (1.62\%).
Table~\ref{tab:factors} summarizes the factors entering into the
$B(K_L^0 \rightarrow \mu^\pm e^\mp)$ upper limit calculation.

\begin{table}
\begin{center}
\begin{tabular}{@{\hspace{2pt}}l@{\hspace{5pt}}r@{\hspace{2pt}$\pm$\hspace{2pt}}l@{\hspace{2pt}}|
@{\hspace{5pt}}l@{\hspace{5pt}}r@{\hspace{2pt}$\pm$\hspace{2pt}}l@{\hspace{2pt}}}
\hline
\hline
$B_{\pi \pi}$  & 0.002067 & 0.000035 &
$A_{\mu e}$ & 1.14\% & 0.006\% \\
$f_{\pi \pi}$ & 0.959 & 0.0058 &
$\epsilon^{\mathrm{L1}}_{\mu e}$ & 0.974 & 0.0046 \\
$R$ & \multicolumn{2}{@{\hspace{13pt}}c|@{\hspace{5pt}}}{$2 \times 10^4$} &
$\epsilon^{\mathrm{L3}}_{\mu e}$ & 0.936 & 0.0071 \\
$N_{\pi \pi}$ & 79,089 & 379 &
$\epsilon^{\mathrm{PID}}_{\pi \pi}$ &  0.978 & 0.0024 \\
$A_{\pi \pi}$ & 1.62\% & 0.007\% &
$\epsilon^{\mathrm{PID}}_{\mu e}$ & 0.928 & 0.0045 \\
\hline
\hline
\end{tabular}
\caption{Factors in the calculation of the $B(K^0_L \rightarrow \mu^\pm e^\mp)$
upper limit.}
\label{tab:factors}
\end{center}
\end{table}

The resulting 90\% confidence level upper limit on the branching fraction is 
$B(K^0_L \rightarrow \mu^\pm e^\mp) < 4.7\times10^{-12}$.
This represents the most sensitive search for $K^0_L \rightarrow \mu^\pm e^\mp$
to date.

We acknowledge the support of the BNL staff, particularly
H.~Brown, R.~Brown, R.~Callister, A.~Esper,
F.~Kobasiuk, W.~Leonhardt, M.~Howard, J.~Negrin, and J.~Scaduto. 
R.~Atmur, K.~M.~Ecklund, M.~Hamela, S.~Kettell, D.~Ouimette, B.~Ware,
and S.~Worm played key roles in 
the design and fabrication of important parts of the experiment.
C.~Allen, G.~Bowden, P.~deCecco, P.~Coffey, M.~Diwan,
M.~Marcin, C.~Nguyen, 
A.~Schwartz, and E.~Wolin
 made important contributions during the early
phases of the experiment. We thank V.~Abadjev,  P.~Gill, N.~Mar,
J.~Meo, M.~Roehrig, and M.~Witkowski 
for valuable technical assistance.
We also thank the SLAC Computing Division and the
BNL CCD for assistance with data processing. 
This work was supported in
part by the U.S. Department of Energy, the National Science Foundation,
the Robert A. Welch Foundation, and Research Corporation.


\begin{thebibliography}{16}
\bibitem{ptin} A correlated cut between $p_T^{^{_\parallel}}$ and $p_e$
was made: $p_T^{^{_\parallel}}<(3 \, p_e - 2.5) \times 10^{-3}$, where
$p_T^{^{_\parallel}}$ and $p_e$ are in GeV/c.
\end{thebibliography}
\end{document}